\documentclass[a4paper,fleqn]{mnras}

\usepackage[T1]{fontenc}
\usepackage{ae,aecompl}

\usepackage{graphicx}   
\usepackage{amsmath}    
\usepackage{amssymb}    
\usepackage{longtable}

\title[Transients of AD 668 and 891 were comets]{``Novae, supernovae, or something else?'' -- 
(Super-)nova {\it highlights} \\ from Hoffmann \& Vogt are quite certainly comets (AD 668 and 891)}

\author[R. Neuh\"auser, D.L. Neuh\"auser, J. Chapman]{
Ralph Neuh\"auser,$^{1}$\thanks{E-mail: ralph.neuhaeuser@uni-jena.de}
Dagmar L. Neuh\"auser,$^{2}$
Jesse Chapman$^{3}$
\\
$^{1}$Astrophysikalisches Institut und Universit\"ats-Sternwarte, Friedrich-Schiller-Universit\"at Jena, 
Schillerg\"a\ss{}chen 2-3, 07745 Jena, Germany  \\  
$^{2}$Independent scholar, 39012 Meran/o, Autonome Provinz Bozen, S\"udtirol/Provincia Autonoma di Bolzano, Alto Adige, Italy \\
$^{3}$Department of East Asian Studies, New York University, New York NY, 10003, United States
}

\date{Accepted 2020 October 26. Received 2020 October 26; in original form 2020 September 21}

\pubyear{2020}

\begin{document}
\label{firstpage}
\pagerange{\pageref{firstpage}--\pageref{lastpage}}
\maketitle

\begin{abstract}
Galactic novae and supernovae can be studied by utilizing historical 
observations, yielding explosion time, location on sky~etc. Recent publications 
by Hoffmann \& Vogt present CVs, supernova remnants, planetary nebulae etc. as 
potential counterparts based on their list of historically reported transients from the Classical Chinese text corpus. 
Since their candidate selection neglects the state-of-the-art (e.g. Stephenson \& Green), and 
since it includes `broom stars' and `fuzzy stars', i.e. probable comets, we investigate their catalogue 
in more detail. We discuss here their two highlights, the 
suggestion of two `broom star' records dated AD 667 and 668 
as one historical supernova and of the `guest star' of AD 891 
as recurrent nova U Sco. 
The proposed positional search areas are not justified due to translation and dating
problems, source omission, as well as misunderstandings of historical Chinese astronomy and unfounded textual interpretations. 
All sources together provide 
strong evidence for comet sightings in both AD 668 and 891 -- e.g., 
there are no arguments for stationarity. The AD 667 record 
is a misdated doublet of 668. Our critique pertains more
generally to their whole catalogue of `24 most promising events': 
their speculations on counterparts lack a solid foundation and should not be used in follow-ups.
\end{abstract}




\section{Introduction}

There are many historical records about transients, 
which can include supernovae (SN), novae, comets, variable stars,
but also aurorae, meteors, bolides, etc.
Since it is difficult to distinguish between such phenomena, one has to set criteria
(e.g., Stephenson \& Green 2005; R. Neuh\"auser \& D.L. Neuh\"auser 2015):
while novae and SNe are stationary 
and star-like, comets move relative to the stars and often show tails -- 
and this has to become clear from the transmitted sources
(e.g., D.L. Neuh\"auser et al. 2020).
For application to modern astrophysical problems, one has to follow a reliable methodology:
literal translations based on critical text editions, possibly dating corrections, context,
parallel sources, further transmissions, close reading etc. (R. Neuh\"auser, D.L. Neuh\"auser \& Posch 2020).

Most recently, by searching for historical nova eruptions,
Hoffmann, Vogt \& Protte (2020, henceforth HVP20) present a list
of `24 most promising events' based on parts of the Classical Chinese text corpus.
They claim to have used a new method, e.g. re-interpretating the texts
and excluding comets and meteors.
Astrophysical objects in their positional error circles, like CVs, are then
discussed in Hoffmann \& Vogt (2020a, henceforth HV20a)
and Hoffmann \& Vogt (2020b, henceforth HV20b).
As their main highlights, they feature: \\
(i) the `broom star' records for AD 667 and 668 together would be a SN seen for one year
(and they suggest a SN remnant), and \\
(ii) the `guest star' of AD 891 would be recurrent nova U Sco. \\
We present strong evidence that the records are parts of comet occurrences --
the state-of-the-art is not considered in HVP20, HV20a,b, 
where published evidence that contradicts their claims is missing. 

Historical records from East Asian court astronomers of the last two millennia use their
nomenclature, calendar, and celestial system to specify details regarding each records, 
for instance observing time and location on sky;
background is provided, e.g., in Sun \& Kistemaker (1997, henceforth SK97);
philological and cultural competence is needed (Chapman, Csikszentmihalyi \& Neuh\"auser 2014;
Chapman et al. 2015). 

A recent review on historical SNe is found in Stephenson \& Green (2002);
a few additional Arabic language records of SNe 1006, 1572, and 1604 are collected in
Rada \& R. Neuh\"auser (2015), R. Neuh\"auser et al. (2016, 2017a), and R. Neuh\"auser, Ehrig-Eggert \& Kunitzsch (2017b).
Recent work on historical `guest stars' including novae and SNe can be found in Stephenson \& Green (2005, 2009), 
a cometography in Kronk (1999), a critical collection of Chinese records of
`broom' and `fuzzy stars' in Pankenier, Xu \& Jiang (2008, henceforth P+08) --
all ignored in HVP20 and HV20a,b. 
The `24 most promising events' in HVP20 include transients called `broom star' or `fuzzy star',
and/or were rejected previously for other reasons;
we investigate as examples their two main highlights:
the Japanese `guest star' record of AD 891 is not about recurrent nova U Sco,
but most probably about the last position of a well-known comet also observed in China, Europe, and Arabia (Sect. 2),
and the `broom star' records of AD 667 and 668 together are not one
new historical SN, but a comet with tail and motion relative to the stars in AD 668 (AD 667 misdated).
We finish with a brief summary.

\section{The Japanese `guest star' record in AD 891: not nova U Sco, but most likely the last sighting of Alfred's comet}
 
The interpretation of the AD 891 guest star in HVP20 and HV20a,b is based purely on the record 
from Japan as translated in Xu, Jiang \& Pankenier (2000, pp. 136, 332), additions in square brackets: \\ 
891 May 11: `3rd year of the Kampy\=o reign period, 3rd month, 29th day [of lunar month], [day] {\it jimao} [16]. 
At the hour of {\it hai} [21-23h], there was a guest star [{\it ke xing}] east of 
the star Dongxian at a distance of about one {\it cun}' 
(Mei getsu ki ch.\,23, compiled by Fujiwara Sadaie AD 1162-1241,
identical in Nihon kiryaku zen 21). 

Classical Chinese does not distinguish between singular or plural, e.g., `star' or `stars'; 
the four stars of Dongxian built the eastern wall of an enclosed market, 
most likely $\phi, \chi, \psi, \omega$ Oph (SK97, pp. 72, 150),
which one was meant is unspecified. The linear scales {\it cun} (also {\it chi} and {\it zhang}) 
were applied as angles to tail length and angular separations between celestial objects 
(e.g., the last position of 1P/Halley in AD 760, Xu et al. 2000; D.L. Neuh\"auser et al. 2020). 
The ganzhi date {\it jimao} is from the 60-day (sexagenary) cycle, 
an alternative to the day within the lunar month, here `29th day'. 
 
HVP20 `selected those [old events] ... not only visible within a certain hour (to exclude meteors)' (abstract)
-- while the small asterism Dong\-xian is seen all night on May 11, the phenomenon is reported for 21-23h
in the two Japanese records in Xu et al. (incorrect in HVP20 table 3);
this record should be excluded, but it is listed as one of their `24 most promising events' 
for novae etc. (HVP20; HV20a,b).\footnote{Also 
the event for 1497 Sep 20 from China contradicts their own criterion, 
reported just `at dusk' (Xu et al. 2000, p. 143), although the position would be visible all night. 
The given `$\beta$ UMi' as alleged identification of `Tianji star' (HVP20, table 3) demonstrates that what was 
announced as their own `new method', namely `a careful re-reading of the respective historical records and 
an independent interpretation of the positions given therein (with regard to the existing works on the 
identification of old ancient names)' (HVP20), was not applied: `Tianji' is in the translation 
of Xu et al. (2000), a typo for `Tianjiu', see Chinese text on their pp. 336/337 (and Stephenson \& Green 2009). 
`Tianjiu' is an asterism of ten stars with 22 And (SK97, p. 191) and/or $\theta$ And (P+08, p. 462), 
ancient maps confirm the number (Stephenson \& Green 2009: `Tianjiu consists of the three stars $\theta, \rho, \sigma$ And', 
which is incomplete). Three asterisms with different Chinese characters are all transcribed `Tianji', 
none of them with $\beta$ UMi, which is one of the five stars of Beiji -- a prominent asterism near the North Pole (SK97, pp. 217ff).}
Since HVP20 use only this source (without duration, stationarity is questionable), 
it {\it seems} to be a brief, unspecified transient event. 
However, HV20b suggest recurrent nova U Sco as its counterpart. 
Furthermore, in their table 3 regarding AD 891, HVP20 claim to `consider only stars in the eastern half' -- 
U Sco is {\it west} of the Dong\-xian skeleton (HV20a figure A-16); neither their assumption that the `principle star' 
of Dongxian was meant (it is disputed, which star was the principle star, see SK97, p. 48) 
nor the size of their positional error circle of $3^{\circ}$ (HVP20 table 3) is justified. 
While the source says 1 {\it cun} ($\sim \, 0.1^{\circ}$) east of `an unspecified member' 
(Stephenson \& Green 2009, p. 44), HV20b `postulate a corrupted text preserving 1 cun instead of 1 chin ($\sim 1^{\circ}$)' 
(section 5.3) -- they mean `chi'. 
In sum, only the textual (against their own selection criterion) and positional interpretation seems to be corrupt. 
As we will see next, further historical records and scholarly research 
with evidence against their claim is not considered; they do not discuss alternatives, 
e.g., planet Uranus ($\sim 5.5$ mag) inside their error circle on the given date -- 
however, another option is more reliable. 

P+08 (pp. 102, 535-6) list more records from China (a and b); text~b is already in Ho (1962)
no. 313 together with the `guest star' report from Japan (May 11),
with a similar translation as above -- Ho is used in HVP20 and HV20a,b, but not here; AD 891 May 12: \\
(a) `2nd year of the Dashun reign period of Emperor Zhaozong of the Tang Dynasty, 4th month, 
day {\it gengchen} [17]; a broom star [{\it hui xing}] entered Taiwei' (Xin Tang shu: Zhaozong ji, AD 1061). \\
(b) `[same date] there was a broom star [{\it hui xing}] over 10 {\it zhang} long in Santai that travelled eastward 
and entered Taiwei, sweeping Dajiao and Tianshi. 
In the 5th month, day {\it jiaxu} [11 = July 5], it no longer appeared [lit.: was not seen]' (Xin Tang shu: tianwen zhi).  

A {\it hui xing} is a star ({\it xing}) with a tail like a broom ({\it hui}), 
the common wording for comets: probably, the longest tail length here was `over 10 {\it zhang}' ($\sim 100^{\circ}$).
These reports are transmitted in the Xin Tang shu, the `New history of the Tang dynasty'. 
Text (a) is shortened, the comet was discovered in the night May 12/13 presumably when entering the enclosure Taiwei, 
but (b) gives more information on the comet's path: it appeared at Santai ($\iota, \kappa, \lambda, \mu, \nu, \xi$ UMa) 
moving east to the Taiwei enclosure (stars in CrB, Leo, Vir), then passing Dajiao ($\alpha$ Boo) 
and the Tianshi enclosure (stars in Her, Ser, Oph, Aql); applying source critique shows
that text (a) combined the first date with a later position -- 
the transmitted records are compilations, which shortened significantly most 
of the original protocols.

Stephenson \& Green (2009) thought that `it seems likely' that the reports from Japan and China pertain to 
two different objects, because `on May 12, the broom star was at Santai, some 90 deg from Dongxian' (p. 44); 
however, let us consider the comet path: Dong\-xian in Oph is just off the south gate of the Tianshi enclosure; 
hence, the position of the `guest star' (Japan) fits well with the last reported area for the `broom star' (China), 
Tianshi, close to the ecliptic, a location typical for departing comets (see Ho 1962, no. 313; Kronk 1999). 
The timing also fits: the comet record from China (b) gives `not seen' on July 5, 
that could be due to clouds, faintness, or disappearance of the comet; 
either way, during July Dongxian is seen from Kyoto best `at the hour of {\it hai}' (21-23h), 
in the previous double hour the Sun is setting, afterwards the asterism sets. 
Thus, most likely, the Japanese compilation transmits the position, where the comet vanished -- 
however, recorded under the date of first sighting, similar to text (a) from China. 

Weather may have influenced the observational records from China;
a misdated doublet of the Chinese transmission to AD 893 
(P+08 (b), pp. 103, 536, also Ho 1962 no. 315, the title of the reign period is confused)
gives: `then [the broom star] was not seen due to clouds and overcast' [condition]. 
A bad weather period is also reported before the comet appeared 
(additionally to P+08 for AD 893 (b), our translation): 
`Year 2, 2nd month: the sky was overcast for a long time ({\it jiu yin}). 
Upon the {\it yiyou} [22] night of the 4th month [converted to AD 891: May 17/18], 
the clouds opened some. 
There was a comet in Shangtai [upper step of Santai] ...' (then like above). 
In this misdated transmission, which could come from a different location, 
the comet seems to be detected about May 17/18. Hasegawa (2002) 
calculated approximate orbital elements for AD 893 --  
but the ganzhi dates are only consistent for AD 891: 
there would be no `day jiaxu [11] in the 5th month' in AD 893 (see source b from AD 891).

The appearance of the impressive comet is also attested in Europe and Arabia
(e.g., Kronk 1999), not noticed by HVP20 and HV20a,b,
a reliable source is the Anglo Saxon Chronicle (England): \\
`AD 891 ... And the same year after Easter [Apr 4], at the rogation days [May 10-12] or before, 
there appeared the star which is called in Latin cometa. 
Some men say that it is in English the long-haired star, for there shines a long ray from it, sometimes on one side, 
sometimes on every side' (Whitelock 1979, p. 201). 

From Arabia, Ibn al-Jawz\textit{\={\i}} (al-Muntazam f\textit{\={\i}} al-t\=ar\textit{\={\i}}kh, AD 1201) reported:
`AH 278 (AD 891): A very brilliant star rose on 28 Mu\d{h}arram [May $13 \pm 2$] 
and then its brilliance became locks of hair' (Cook 1999); 
similar by Ibn al-Ath\textit{\={\i}}r (al-K\=amil f\textit{\={\i}} al-t\=ar\textit{\={\i}}kh). 

The discovery might have been in England; with full historical right, Schove (1984) 
called this long-haired star `Alfred's comet' after the influential King of Wessex and then of all Anglo-Saxons (AD~848/9-899), 
who initiated the Chronicle compilation;
designation in Kronk (1999): comet X/891~J1.
Sometimes the comet is dated AD~892, which is not justified -- 
a contemporary source clarifies the year: `AD~891. Stella cometis. Eclypsis solis', 
first the comet, then the solar eclipse of 891 Aug 8 
(Ann. Alamannic. cont. Sangallensis tertia, St. Gallen, AD 926, MGH SS 1, p. 52, ed. Pertz). 

In sum, caution is needed when considering historical observations of transients as evidence for 
rare high-energy events like novae or SNe: similar to AD 891, records of `guest stars' in AD 85 and 1166 
pertain most likely to `broom star' sightings, for a `guest star' in AD 64, among others, 
a cometary interpretation is more reliable (Stephenson \& Green 2009) -- 
the latter is also one of the `most promising [non-cometary] events' in HVP20. 
In P+08 many such conversions are attested, but already in Ho (1962) the problem was well-known
(e.g., no. 228, regarded as nova by others): 
`we shall come across several cases where guest stars turned into ({\it hui}) comets and vice versa' (p. 137). 
Comets can be observed as small, large, or so-called fuzzy (guest) stars, 
in particular during the first and last sighting (see also the records from England and Arabia)
-- these phenotypical descriptions do not contradict their physical nature. 

\section{The `broom star' records for AD 668 and 667: neither a new supernova in AD 667-8 nor \\~~~~Nova V392 Per, but most certainly a comet}

HV20a present the event from AD 668 as nova. 
HV20b interpret the `broom star' records for AD 667 and 668 together as historical SN seen for one year. 
Since the reported phenomena are explicitly called {\it hui xing} (lit. `broom star', star with tail),
they are not listed in Xu, Jiang \& Pankenier (2000), a compilation in particular of likely stationary {\it ke xing} (lit. `guest star'). 
The AD 667 and 668 reports are instead found in the new collection of Chinese records of
`broom' and `fuzzy stars' as well as likely non-stationary guest stars (Pankenier, Xu \& Jiang 2008). 
HVP20 use the old, partial translation from Ho (1962), a collection of 579 `comets and novae' -- 
completely revised by Xu et al. (2000) and P+08 with classical Chinese texts. 
We will see that the AD 667 text is a misdated copy of the AD 668 sources 
and that the transmissions can only be interpreted as a single comet. 

We present step by step the transmitted texts from China and Korea as translated in P+08 (pp. 73-74, 518-519), 
variants from Ho (1962, no. 251, 252, but in Pinyin transcription) in round brackets. \\
First, 668 May 17\,-\,June 14, i.e. the full 4th lunar month, P+08: \\
(c) `1st year of the Zongzhang reign period of Emperor Gaozong of the Tang Dynasty, 4th month; 
a broom [{\it hui}] star appeared [or: was seen] at (Ho no. 252: above) Wuche ... 
on the 22nd day, the star [{\it xing}] was extinguished' 
(Jiu Tang shu 36.1320: tianwen zhi, compiled AD 945, for text omission, see below).

This short notice (as given by Ho 1962, no. 252) about a broom star in the `Old history of the Tang Dynasty' 
is the basis for the interpretation as recurrent nova in HV20a and (together
with the misdated doublet in 667, Ho no. 251) as SN in HV20b.
The asterism Wuche ($\alpha$ Aur + 4 stars) is mainly the same as the western 
constellation figure Auriga; Ho (1962, no. 252) translated erroneously 
`above Wuche'.\footnote{The preposition {\it shang}, i.e. `above', does not belong
to `Wuche': after the Chinese text `hui jian Wuche' (`broom [star] appeared [at] Wuche'), 
text~(c) has `shang bi zheng dian' for `the Emperor avoided the main hall' -- 
`shang' belongs to the following clause and refers to the emperor;
the separation between {\it shang} and the preceding clause is further clarified in text~(a) below, 
where the ganzhi date {\it yichou} is in between `Wuche' and `shang';
another parallel text (b) has {\it yu} for `at' (Wuche).} 
Thus, the positional interpretation by HVP20 (table 2) is not justified from the Chinese transmission: 
the alleged `above' is interpreted for sightings during `evening' or `morning', 
leading to two error circles (with questionable different radii, HVP20 table 3 and 
HV20a figure A-12).
However, the text also has `in the northeast', omitted by Ho (1962), see below, 
so that the observation was obviously performed at the end of the night.

Although HV20a (section 4) `believe' that an `earlier eruption was detected by ancient Far Eastern observers',
their `excellent candidate for a recurrent nova', namely V392 Per (also given as S1065, but in fact referring to S10653), 
is not found in their figure A-12: in table 2, it is stated that V392 Per was `not visible at twilight', 
while in section~4, they wrote that `the historical nova ... was only visible at twilight', 
yet `a coincidence with a possible nova in 668 CE is not completely excluded'. 
At the end of the reported range of dates (`22nd day' is June 6/7), the position of their unjustified search circle 
could be seen before astronomical twilight (already $\sim 10^{\circ}$ above horizon). 
HVP20 have a question mark for the duration in table 3, but it is clearly given, see source (a) next.

Further references are mentioned already in Ho 1962 (no. 252),  
but parallel transmissions were not accounted for by HVP20, HV20a,b -- 
P+08 give additional sources, both 668 May 18: \\ 
(b) `1st year of the Zongzhang reign period of Emperor Gaozong of the Tang Dynasty, 4th month, day {\it bingchen} [53]; 
[there was] a broom star [{\it hui xing}] [that] emerged at [or: in ({\it yu})] Wuche ...' 
(Xin Tang shu 3.36-67: Gaozong ji, AD 1061, omissions below). \\
(a) `1st year of the Zongzhang reign period of Emperor Gaozong of the Tang Dynasty, [summer,]
4th month, day {\it bingchen} [53]; [there was] a broom star [{\it hui xing}] [that] appeared [or: was seen] between 
[{\it zhi jian}, lit.: in the space of] Bi and Mao. ... 
on day {\it yihai} [12 = June 6], the broom star was extinguished [or: vanished]' 
(Jiu Tang shu 5.91-92: Gaozong, AD 945, omissions below).

All texts are consistent regarding the duration: while (c) has only the lunar month, (a) and (b) give a date, 
4th month, ganzhi day {\it bingchen} [53], i.e. night May 18/19, for first detection (not given in HVP20, HV20a,b);
disappearance is stated in (c) for the 22nd day of the 4th lunar month, which is ganzhi day {\it yihai} [12] 
in the 60-day-cycle, recorded in (a), the night June 6/7 (19 nights).

The `space of Bi and Mao' cover right ascension ranges, so-called lunar mansions (LM): 
Mao is LM 18 from 7 or 17 Tau to $\epsilon$ Tau, Bi is LM 19 from $\epsilon$ Tau to $\phi$ or $\lambda$ Ori 
(e.g. SK97, p. 52). 
Mao and Bi are also asterisms, Mao is identical with the Pleiades (7 stars), Bi with $\alpha$~Tau, the Hyades, etc. (in total 8 stars). 
However, the wording {\it jian} in combination with such entities (here: Bi and Mao) indicates lunar mansion space, 
which often feature in comet reports (e.g. 1P/Halley AD 760 and 837, e.g. Xu et al. 2000). 
We have now various information about the `broom star': 
first seen at the asterism Wuche, which lies (almost entirely) in LM Bi, but not in LM Mao; 
and, also, since LM Bi (19) is reported before LM Mao (18), probably not just by chance, 
we may conclude that the comet was flying from East to West (LM 19 to 18). (Even 
if {\it jian} did point to `between the asterisms Bi and Mao', this position would be
disjunct from `at Wuche', and hence showing motion.)
Thus, the observations were at the end of the night -- the comet had just passed the Sun.
 
This scenario is supported by a further source from Korea (see P+08 (a), pp. 74, 519), 
interestingly also given under Ho no. 252, but not taken into account by HVP20 nor HV20a,b. \\
668 May 17\,-\,June 14, 4th lunar month, P+08 (a): 
`8th year of King Munmu of Silla, 4th month; a broom star 
guarded [{\it shou}] Tianchuan' (Samguk sagi, Jeungho munheon bigo); text (b) in footnote 5.

Although this Korean chronicle is sometimes a few years off, date and position fit well to the path given 
in China: since the eastern end of the asterism Tianchuan 
(from west to east along $\eta, \gamma, \alpha, \delta, \mu$ Per, in total 9 stars) 
is situated in LM Mao, and the verb {\it shou}, i.e. guard or linger, 
means to `remain stationary within 1 to 2 du [$1-2^{\circ}$] of another body' (P+08, p. 467), 
the report seems to give information where the `broom star' slowed down 
and was seen last. 

Next, we will discuss the text omitted in the above quotations (a) to (c) from China. 
Although the positions for the `broom star' differ, (a) has `in the space of Bi and Mao', (b) and (c) `at Wuche', 
the omitted context in (a) and (c) is almost identical, (b) is very shortened -- 
this shows that these transmissions refer to the same `broom star'. 
Parts of the omissions are given in P+08 (c), but only (a) has a further date, so we choose it for our translation (from JTS): \\
`[Day] {\it yichou} [2 = May 27], the emperor avoided the main hall and reduced [his consumption of] rich foods. 
[He] summoned ... groups of officials [asking] each to submit a memorial, ... 
their words: ``When the Star became fuzzy [{\it xing bo}]\footnote{P+08: `the star subsequently became fuzzy', but 
`subsequently' is implied, there is no character for it in the Chinese text.}, 
its bright rays [{\it guang mang}, lit. also: light thorn/s] small/er, 
this is not an ill-omen for the state and is insufficient [cause] for his majesty to labour [himself with] sagely worries. 
Please sit [in] the main hall, restore the normal meals.'' 
The Emperor said: ``I received [the duty to] offer [sacrifices] ...  
[When a sign of] blame is seen in the heavens, it admonished [me for] my lack of virtue ...'' 
The many ministers again advanced saying (text c names Xu Jingzong, AD 592-672): 
``The star became fuzzy [{\it xing bo}] in the northeast (text c adds: the royal troups denounce the guilty)
-- this is a sign that Goryeo is about to be destroyed.''
The Emperor said: ``The people of Goryeo are my people. When acting as the lord of the myriad nations, how
can I push [my] errors over a little hedge.'' 
In the end, he did not follow the requests made of him.'

On day {\it yichou} [2], May 27, the experts brief the emperor that the broom star (detected on May 18/19), 
which they might have observed the nights before `in the northeast', now `became fuzzy' and `its bright rays small/er': 
evidently, the tail of the comet (`broom') had lost already much of its length and/or brightness, 
the head of the comet (`star') had only short, weak rays.\footnote{P+08, p. 6: 
`In the past, some scholars have been perplexed by the compound {\it xingbo}, which appears to defy Chinese grammatical 
conventions by having {\it xing} `star/celestial body' modify bo `be fuzzy/bristle'. 
However, {\it bo} has a verbal sense here, meaning `to become fuzzy or bushy'. 
This is entirely consistent with cometary records where it is generally used to describe the appearance of 
tail-less comets or the changed aspect ... Rendering {\it xing bo} as ``bushy [or fuzzy] star'' ... is misleading 
in that it obscures the possibility that {\it bo} may imply a change of appearance' 
(P+08), see also Ho (1962, pp. 136-7) and Stephenson \& Green (2009, p. 31). 
The `24 most promising [non-cometary] events' in HVP20 contain at least five `broom stars' 
plus five more which are given as `fuzzy stars'; 
the terms `broom' and `fuzzy' are neglected in HVP20 (section 2.1) for discriminating
between phenomena.}
This change of getting smaller and fainter is interpreted to signal China's loss 
of the kingdom Goryeo in AD 668 thereby uniting the Korean peninsula under the Silla dynasty.\footnote{P+08 
text (b) from Korea (very much shortened from the Chinese sources a and c) in the chronicle of the
kingdom Goryeo:
`27th year of King Bojang (r. AD 642-668, d. 682) of Goguryeo [=Goryeo], summer, 4th month (668 May 17\,-\,June 14); 
a broom star was seen between Bi and Mao' from Jeungho munheon bigo, which continues (our translation): 
`In Tang (i.e. China), Xu Jingzong said: ``a broom (sic) appears [in the] northeast, it is a sign
that Goguryeo is about to be destroyed''\,'.}

It is beyond the scope of this paper to survey further transmissions from Europe etc.; 
it is known that Theophanes reported a `sign' at this time (Schove 1984) --
unfortunately, it was not very specific, but comets are often reported this way.

In sum, the transmitted details from East Asia fulfill all five comet criteria which help to identify a likely 
true comet (D.L. Neuh\"auser et al. 2020) -- with full right, the `broom star' is listed in various 
catalogues of comets (e.g., Kronk 1999): \\ 
(i) Timing: observed at night-time or twilight -- is given. \\ 
(ii) Position of first and/or last sighting: often close to Sun and/or near ecliptic -- is given. \\ 
(iii) Colour and form: the phenotypical description as {\it hui xing} indicates a {\it star} with a tail (`broom'), 
whose change (`became fuzzy' etc.) is also mentioned. \\
(iv) Motion relative to stars: detection in/at Wuche (Aur) and probably last sighting at Tianchuan (Per), 
together with probably first in LM Bi (19), then in LM Mao (18), the comet path is given. \\
(v) duration: 19 nights -- consistent with comets (if `extinguished' ({\it mie}, texts a and c) means that it
was already too faint June 6/7). 

As mentioned, HVP20 and HV20a,b use only the brief text from source (c), as in Ho (1962) no. 252, 
quoted at the beginning of this section,
with an erroneous translation (`above Wuche')
as well as presumably questionable duration (HVP20, table 3) or unclear dating 
(`only several days (maximum three weeks)' in HV20b section 5.3, but see also their footnote 7). 
HV20b argue that the phenomenon was `only visible in twilight' (not correct) and, hence, `probably rather bright', 
which would `make[s] the observation of a SN more likely' than a nova\footnote{Nota bene: A very bright nova or SN appearing 
extended by strong scintillation (therefore called `fuzzy') can be an acceptable alternative 
only in exceptions like day-time sighting and explicit mention of stationarity --
however, even SNe 1006 and 1054 were not described as `fuzzy'.} -- 
their `excellent candidate for a recurrent nova', namely V392 Per (HV20a), 
is not mentioned any more in the summarizing results (HV20b, table 9). 
Yet, HV20b state that the short duration `would be atypical' for a SN -- a comet (sic) or nova would be more likely. 
However, then HV20b suggests `a new hypothesis', because `something else appears really strange': 
`In two subsequent years, there are two comets (or novae) at almost the same position in the sky 
(close to the Wuche-asterism)'; HV20b `consider it more likely that the two records together report 
one supernova than that there were two comets at almost the same position'.

Let us take a look at the text from China from the presumably preceeding year,
which is meant here -- as translated in P+08 (pp. 73, 518),
variants from Ho (1962, no. 251) in round brackets: \\
AD 667 May 24: `2nd year of Qianfeng reign-period of Emperor Gaozong Tang Dynasty, 4th month, day {\it bingchen} [53]. 
There was a broom star [{\it hui xing}] in the northeast, between (Ho: among) Wuche, Bi [LM 19], and Mao [LM 18] 
[lit.: it was located in/at [{\it zai}] Wuche, in the space [{\it jian}] Bi, Mao]; 
by day {\it yihai} [12], it did not appear [or: was not seen]' (Xin Tang shu 32.837, Tianwen zhi).

This shortened transmission in the `New history of the Tang' is likely concatenated from 
the older Jiu Tang shu (texts a and c), or common sources. This procedure is not unusual. 
Compilations work that way, copying errors happen. 
The misdating was already noticed by P+08 (p. 73): 
there was no day {\it yihai}, i.e. day 12 in the 60-day-cycle, 
in the 4th month for the given reign-period, but one year later it did exist.
\footnote{Probably 
the occurrence was first listed in Qianfeng, year 3 rather than Qianfeng, year 2: the reign-period changed 
from Qianfeng to Zongzhang sometime in the third month of AD 668 (April 17\,-\,May 16).
If the record was originally dated to Qianfeng, year 3, a copyist might not have realized that Qianfeng 
had ended by the fourth month, and, rather than correctly changing the reign period, incorrectly changed the year.}

Since all other information given -- name of emperor, day numbers, lunar month, `broom star', position, path, 
duration (again 19 days, in HVP20 table 3 only `18') -- are identical, one can conclude that the text of AD 667 
belongs clearly to the transmitted sources for AD 668. In HVP20, this misdated doublet is listed as one of 
the `24 most promising events'. In HV20b, table 9, the best results are summarized: 
for AD 667 `appearance of a SN'; for AD 668 `disappearance of SN 667 
(consider SNR G160.9+02.6 (?))'.
In the `discussion of individual events' (HV20b, section 5.3), their `new hypothesis' postulates the 
need to `re-interpret the position' for AD 667: `it was originally not meant 
to have occurred between three asterisms (which would be, indeed, a very unusual description) 
but next to the Wuche asterism (as event 668)', HV20b mean their wrongly deduced position (`above Wuche'). 
Thus, the `two separate sightings' are fabricated to pertain to `one supernova observation 
from May 667 to June 668 CE' (HV20b, abstract) -- but in fact almost none of the six criteria for reliable 
historical SNe (given in Stephenson \& Green 2005, pp. 218-219) are fulfilled.\footnote{`Long duration' (i), 
`fixed location' (ii), `low Galactic latitude' (iii), `no evidence of significant angular extent' (iv), 
and `unusual brilliance' (v) are construed for AD 667-8 in the overinterpretation by HVP20 and HV20a,b,
even (iii) is not that certain regarding the true scenario,
and `independent records' (vi) are absent.}

The unfounded speculations (one more column in HV20b section~5.3 is filled with SNR 
considerations)\footnote{Earlier, Xi \& Po (1966) suggested as SNR counterpart 
for the AD 668 `broom star' the radio source CTB-13, based on partly incomplete sources 
and questionable translation, but the record for AD 667 is already omitted;
Chu (1968) on AD 667 and 668: `Since these constellations cover a wide area ... it hardly seems
to be the same localized event', observed `when the Sun is near to these constellations',
i.e. `probably a comet' in May 668. 
These facts are not mentioned by HVP20 etc., but HV20a do cite both otherwise.
The AD 667 record is not listed in Stephenson (1976) on novae and SNe.
For the records AD 667 and 668, Kronk (1999) has one comet in AD 668.} 
in HVP20 and HV20a,b can be seen as an example for how transmissions about transients should not be studied: 
their approach runs the risk of treating historical records as quarry, thereby introducing
unwarranted possibilities into the general astrophysical discourse on historical novae
(see D.L. Neuh\"auser, R. Neuh\"auser \& Chapman 2018, R. Neuh\"auser, D.L. Neuh\"auser, Posch 2020). 


\section{Summary}

As one of two main {\it highlights}, HV20a,b feature a presumable new historical SN from AD 667 to 668
from two `broom star' records in these two years: the former record, however, is a misdated
doublet of the latter, and there are additional Chinese and Korean accounts on the transient object in AD 668
(partly already in Ho 1962, fully in P+08), which provide timing and duration, positions and path,
and that the `broom star' became fuzzy -- all typical for comets.
Their 2nd {\it highlight} is the interpretation of a single brief Japanese account on a `guest star'
in AD 891 as recurrent nova (U Sco), but also here they neglect additional published records
from China, Europe, and Arabia, which together show that the Japanese account 
most likely just gives the last position of Alfred's comet.

In addition, their `24 most promising events' as nova candidates (HVP20) include
several more `broom' and `fuzzy stars'. 
E.g., for the five `broom star' events in HVP20 (tables 2 \& 3), only the texts considered 
by them (from Ho 1962) already fulfil most to all comet criteria (as listed in Sect. 3).
If certain information is missing, e.g. the comet path, one cannot conclude on stationarity.
While HVP20 claim to have eliminated comets
(abstract: `we selected those without movement and without tail (to exclude comets)'),
there are obviously many comets in their `representative shortlist' (of 24 events).
HVP20 do not take into account the state-of-the-art (e.g. Stephenson \& Green 2009 on `guest stars')
and important relevant literature (e.g., Kronk 1999 and P+08 on comets).
Their candidates, positions, and counterparts (HV20a,b) should not be used for astronomical follow-up observations:
even for the few potential nova candidates among their list, positional search fields
by others (e.g., Stephenson \& Green 2009) are more reliable.

\smallskip

{\bf Data Availability.} All data are incorporated into the article.

\vspace{-.5cm}

\section*{Acknowledgements}
We would like to thank the anonymous referee for in-depth analysis, good suggestions,
and encouraging words.


{}





\bsp	
\label{lastpage}
\end{document}